\documentclass[fleqn,twoside]{article}
\usepackage[headings]{espcrc2}
\readRCS
$Id: espcrc2.tex,v 1.2 2004/02/24 11:22:11 spepping Exp $
\usepackage{graphicx}
\usepackage[figuresright]{rotating}

\newcommand{\Dsl}{D\hspace{-2.4mm}/}
\newcommand{\ssl}[1]{{#1}\hspace{-1.8mm}/}
\newcommand{\ave}[1]{\langle {#1} \rangle}
\newcommand{\pb}{\bar q}
\newcommand{\qq}{\ave{\pb q}}

\newcommand{\AmS}{{\protect\the\textfont2
  A\kern-.1667em\lower.5ex\hbox{M}\kern-.125emS}}

\title{Nonlocal quark model beyond mean field and QCD phase transition}

\author{D. Blaschke\address[ITP]{Institute for Theoretical Physics, University of Wroclaw, 50-204 Wroclaw, Poland}%
        \address[BLTP]{Bogoliubov  Laboratory of Theoretical Physics, JINR Dubna, 141980  Dubna, Russia},
        M. Buballa\address[TU]{Institut f\"ur Kernphysik, Technische Universit\"at Darmstadt, D-64289 Darmstadt, Germany},
        A.E.~Radzhabov\address[ISDCT]{Institute for System Dynamics and Control Theory, 664033 Irkutsk, Russia}
        \thanks{Financial support by the grant of Russian President, RFBR (No. 09-02-00749, 09-02-09505) and EMMI.
        Numerical calculations have been carried out on \textit{Blackford MultiCore} computing cluster of ISDCT.}
        and
        M.K.~Volkov\addressmark[BLTP]
}

\runtitle{Nonlocal quark model beyond mean field and QCD phase transition}
\runauthor{D. Blaschke, M. Buballa, A.E.~Radzhabov, M.K.~Volkov}

\begin{document}

\begin{abstract}
A nonlocal chiral quark model is consistently extended beyond mean field using a strict $1/N_c$ expansion scheme. The parameters of the nonlocal model are refitted to the physical values of the pion mass and the weak pion decay constant. The size of the $1/N_c$ correction to the quark condensate is carefully studied in the nonlocal and the usual local Nambu--Jona-Lasinio models. It is found that even the sign of the corrections can be different. This can be attributed to the mesonic cut-off of the local model. It is also found that the $1/N_c$ corrections lead to a lowering of the temperature of the chiral phase transition in comparison with the mean-field result. On the other hand, near the phase transition the $1/N_c$ expansion breaks down and a non-perturbative scheme for the inclusion of mesonic correlations is needed in order to describe the phase transition point.
\end{abstract}

\maketitle
\section{Introduction}

Understanding the QCD phase diagram is one of challenging issues in modern theoretical physics. A description of the most interesting region of phase diagram at low and moderate temperatures/densities requires a non-perturbative approach, which also provides a proper understanding of the chiral quark dynamics and the confinement mechanism.

Until now, the only method which is directly based on QCD and which meets these requirements is lattice gauge theory. Unfortunately, the application of lattice results to experimental data is complicated by the fact that most lattice calculations are performed with rather large quark masses, leading to unphysically large pion masses.

In the present contribution we want to discuss an effective model of low-energy QCD, capable of describing the chiral as well as the deconfinement transitions. As a basis we use the PNJL model, see e.g. \cite{Meisinger:1995ih,Fukushima:2003fw,Megias:2004hj,Roessner:2006xn,Blaschke:2007np}, which generalizes the well-known Nambu--Jona-Lasinio (NJL) model for the chiral quark dynamics by coupling it to the Polyakov loop, which serves as an order parameter of the deconfinement transition.

To get a consistent picture of the hadronic phase it is important to go beyond the mean-field approximation and to include mesonic correlations. In the present work we suggest an improvement of the PNJL model within strict $1/N_c$ expansion scheme and restrict ourselves to the case of zero chemical potential.

\section{Mean field}

The quark sector of the nonlocal chiral quark model is described by the Lagrangian
\begin{eqnarray}
\mathcal{L}_q= \bar{q}(x)(i \Dsl -m_c)q(x)
+\frac{G}{2}[J_\sigma^2(x) + \vec J^{\,2}_\pi(x)] ~,\label{QLagrangian}
\end{eqnarray}
where $m_c$ is the current quark mass, and $D_\mu=\partial_\mu-iA_\mu$ the covariant derivative with a background gluon field $A_\mu \equiv A^a_\mu\frac{\lambda^a}{2} = \delta_{\mu 0} A_0$. The nonlocal quark currents are
\begin{eqnarray}
J_\mathrm{M}(x)&=&\int d^4 (x_1 x_2)f(x_1)f(x_2)\times\nonumber\\
&&\quad\times\bar{q}(x-x_1)\mathbf{\Gamma}_\mathrm{M} q(x+x_2),\label{QCurrents}
\end{eqnarray}
where $\mathbf{\Gamma}_\mathrm{\sigma}= \mathrm{1}$, $\mathbf{\Gamma}_\mathrm{\pi}= i \gamma^5 \tau^a$ with $a=1,2,3$. Spontaneous breaking of chiral symmetry leads to the formation of a quark condensate and generates a dynamical contribution to the quark mass. As a result the Euclidean quark propagator takes the form
\begin{eqnarray}
S_p=(i p\hspace{-1.8mm}/+ m(p^2))^{-1},\, m(p^2)=m_c+m_d f^2(p^2)\nonumber
\label{gap}
\end{eqnarray}
where $f^2(p^2)=\exp(-p^2/\Lambda^2)$ is a (Fourier transformed) Gaussian form factor and $m_d$ is an order parameter for dynamical chiral symmetry breaking. The chiral condensate is obtained from the non-perturbative part of the quark propagator, $S^{np}_p=S_p - S^c_p$, $S^c_p=(i p\hspace{-1.8mm}/ + m_c)^{-1}$.

External vector and axial vector fields can be introduced by a delocalization of quark fields with the help of a Schwinger phase factor. Practically, nonlocal vertices with external currents can be obtained by using rules for derivatives of contour integrals.

\section{Beyond mean field}

\begin{figure}[htb]
\begin{center}
\begin{tabular}{ccc}
        \includegraphics[width=0.10\textwidth]{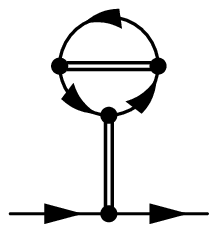}&\hspace{0.1\textwidth}&
        \includegraphics[width=0.15\textwidth]{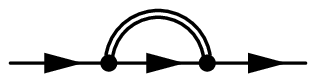}\\
        a && b
\end{tabular}
\end{center}
  \caption{$1/N_c$ corrections to the quark propagator}
    \label{Qucorrection}
\end{figure}

\begin{figure}[htb]
    \centering
        \includegraphics[width=0.4\textwidth]{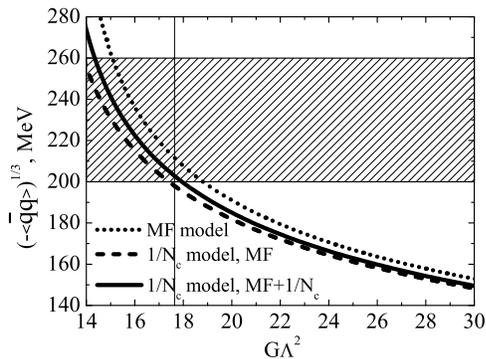}
\caption{Quark condensate in the nonlocal PNJL model in MF approximation (dotted line) and in the $1/N_c$ approximation with parameter adjustment (solid line). The hatched region corresponds to the QCD sum rule limits for the quark condensate: $200 < -\langle \bar{q}q\rangle^{1/3} [{\rm MeV}] < 260$. The vertical line corresponds to the dimensionless coupling for the parameter set No.~4 used in this work.}
    \label{condensate}
\end{figure}

Corrections to the dynamical quark mass beyond the mean-field can be accounted for in a systematic $1/N_c$ expansion scheme \cite{Dmitrasinovic:1995cb,Blaschke:1995gr,Oertel:1999fk} of the quark selfenergy, $\Sigma^{Nc}_p=i\ssl{p}A_p+B_p$ (see Fig. \ref{Qucorrection}), and the quark propagator
\begin{eqnarray}
\left(S^{\mathrm{mf+Nc}}_p\right)^{-1}&=&
S_p^{-1}+\Sigma^{\mathrm{Nc}}_p\nonumber\\
S^{\mathrm{mf+Nc}}_p&\approx&S_p-S_p\Sigma^{\mathrm{Nc}}_p S_p+...
\end{eqnarray}
In order to arrive at a consistent approximation, one needs to take into account $1/N_c$ corrections to the meson propagator, see Ref.~\cite{Oertel:1999fk} for the NJL model and \cite{Plant:2000ty} for its nonlocal generalization. For the present model, we employ a diagrammatic technique developed in \cite{BBRV}. The $1/N_c$ corrections to meson properties will affect the results for the quark condensate via the readjustment of the model parameters ($\Lambda$, $m_c$, $G\Lambda^2$) which are to be chosen such that the physical values for pion mass $M_{\pi^\pm}=139.57$ MeV and weak pion decay constant $f_\pi=92.42$ MeV are obtained at $T=0$, while the dimensionless coupling $G\Lambda^2$ is left as a free parameter. Different parameterizations of the nonlocal model beyond mean field are given in Tab.~\ref{FitModParams}. The corresponding quark condensate is presented in Fig.~\ref{condensate}. We will use the parametrization No.~\textbf{4} for finite T calculations because it provides the highest (pseudo)threshold value of external momentum in quark loop.

\begin{table}[htb]
\caption{Different parameterizations fitted to $M_{\pi^\pm}$ and $f_\pi$. }
\begin{center}
\begin{tabular}{|c|c|c|c|c|c|c|c|c|c|}
\hline
No.&$\Lambda$, MeV & $m_c$, MeV &  $m_d$, MeV  & $G\Lambda^2$   \\
\hline
1         &1479.2         &  2.82          &   139.2         &  13.35         \\
2         & 934.8         &  5.58          &   211.2         &  14.89         \\
3         & 705.9         &  8.64          &   269.1         &  17.06         \\
\textbf{4}&\textbf{670.3} &  \textbf{9.31} &  \textbf{281.9} & \textbf{17.64} \\
5         & 580.5         & 11.78          &   322.5         &   19.72        \\
6         & 500.8         & 14.95          &   373.8         &   22.83        \\
7         & 445.3         & 18.15          &   424.0         &   26.33        \\
8         & 404.4         & 21.37          &   473.4         &   30.20        \\
\hline
\end{tabular}
\end{center}
\label{FitModParams}
\end{table}

In table \ref{FitModParams2} we present the mean-field contributions to pion mass and weak decay constant for different parameter sets together with values for $f_\pi$ estimated from Goldberger--Treiman and Gell-Mann--Oakes--Renner relations. For lower values of the current (and dynamical) quark masses the $1/N_c$ corrections to pion mass and weak pion decay constant amount to 15 MeV and 20 MeV, respectively. For set 4 the corrections are only about 2 MeV and 5 MeV.

\begin{table}[htb]
\caption{Mean-field contributions to $M_\pi$, $f_\pi$ and estimates for $f_\pi$ from low-energy theorems.}
\begin{center}
\begin{tabular}{|c|c|c|c|c|}
\hline
No.&$M_\pi^{\mathrm{MF}}$&
$f_\pi^{\mathrm{MF}}$&
$f_\pi^{\mathrm{GT}}$&
$f_\pi^{\mathrm{GMOR}}$\\

\hline
1         & 155.5          & 72.6         &93.6         &94.7         \\
2         & 144.6          & 83.4         &92.8         &91.3         \\
3         & 142.5          & 87.1         &92.0         &89.9         \\
\textbf{4}& \textbf{142.2} & \textbf{87.6}&\textbf{91.8}&\textbf{89.6}\\
5         & 141.7          & 88.7         &91.4         &88.7         \\
6         & 141.4          & 89.6         &90.9         &87.6         \\
7         & 141.2          & 90.0         &90.6         &86.4         \\
8         & 141.1          & 90.3         &90.4         &85.3         \\
\hline
\end{tabular}
\end{center}
\label{FitModParams2}
\end{table}

Fig.~\ref{condensate} shows that the $1/N_c$ correction to the absolute value of quark condensate is positive for all sets of model parameters. In the local NJL model of Ref.~\cite{Oertel:1999fk} it was found that this correction is negative. However, in the local NJL model due to its non-renormalizability it is
necessary to introduce different regularizations for the pure quark and the meson-quark loops, respectively. In \cite{Oertel:1999fk} a Pauli-Villars regularization has been used for quark loops and a three-dimensional momentum cutoff $\Lambda_M$ for meson-quark loops. In order to study the transition from the local model to a local one let us construct a nonlocal model with three parameters
\begin{enumerate}
\item parameter of nonlocality $\Lambda$
\item parameter of quark loop regularization $\Lambda_q$
\item parameter of meson loop regularization $\Lambda_M$
\end{enumerate}
The local model corresponds to the limit
\begin{eqnarray}
\Lambda\rightarrow\infty~,\quad
\Lambda_q^{}=\Lambda_q^{phys}~,\quad
\Lambda_M^{}=\Lambda_M^{phys}~,
\label{localLimit}
\end{eqnarray}
while the nonlocal model without regularization can be obtained by setting
\begin{eqnarray}
\Lambda=\Lambda^{phys}~,\quad
\Lambda_q^{}\rightarrow\infty~,\quad
\Lambda_M^{}\rightarrow\infty~.
\label{nonlocalLimit}
\end{eqnarray}
For definiteness, let us compare the local model \cite{Oertel:1999fk} with the nonlocal from \cite{Blaschke:2007np} with parametrizations fixed in the MF approximation. Note that for the given parametrizations, the MF quark condensates in the local and the nonlocal model agree within less than $0.5~\%$.

\begin{figure}[!htb]
    \centering
    \includegraphics[width=0.42\textwidth]{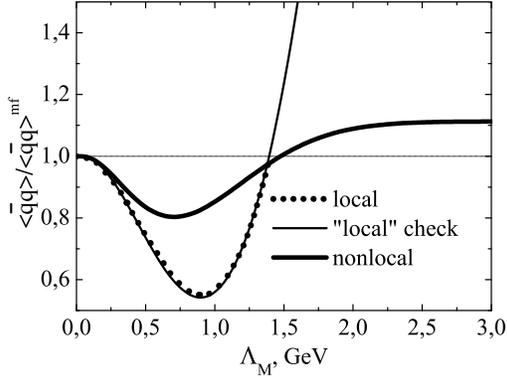}
\vspace{-5mm}
\caption{The ratio $\qq/\qq^{\mathrm{mf}}$ as a function of the meson cutoff $\Lambda_M$. Local results (dotted line) is taken from Fig.3a \cite{Oertel:1999fk}. "Local" check (thin solid line) means the local limit of nonlocal calculations and thick solid line is nonlocal result.}
\label{Ratioqq}
\end{figure}

The next step is to consider the $1/N_c$ corrections and to investigate the role of the mesonic 3D cut-off $\Lambda_M$. For this purpose it is very instructive to study the ratio of the full quark condensate to the MF contribution $\qq/\qq^{\mathrm{mf}}$. In Fig.~\ref{Ratioqq} we compare the $\Lambda_M$ dependence of this ratio for the local NJL model as given in Ref.~\cite{Oertel:1999fk} (dotted line) to that of the nonlocal model (bold solid line) and its local limit (thin solid line). It is very interesting that in the region below ~1.5 GeV these models predict a negative sign for the $1/N_c$ correction whereas for large mesonic cut-off the sign is positive. However, in the nonlocal model the absolute value of the correction saturates for $\Lambda_M$ larger than $\sim 2.5$ GeV, which is  well above actual parametrizations for $\Lambda_q$ and $\Lambda_M$ in \cite{Oertel:1999fk}.

\section{Finite Temperature}

\begin{figure}[htb]
    \centering
        \includegraphics[width=0.42\textwidth]{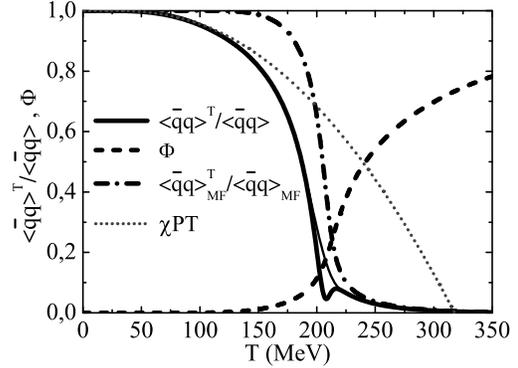}
\vspace{-5mm}
\caption{Temperature dependence of the quark condensate (thick solid line) and the Polyakov loop (thick dashed line) in the nonlocal PNJL model beyond mean field. Thick dash-dotted line: mean field contribution in the $1/N_c$ model; thin dotted line: lowest order chiral perturbation theory ($\chi PT$); thin solid line: na\"ive polynomial interpolation in the unstable region of the $1/N_c$ expansion.}
  \label{DynMF}
\end{figure}

Details of the finite temperature extension of the nonlocal model can be found in \cite{Blaschke:2007np,BBRV}. For the Polyakov loop potential we adopt the logarithmic form of Ref.~\cite{Roessner:2006xn} which has been fitted to the quenched lattice data. In Fig.~\ref{DynMF} we show the resulting temperature dependence of the quark condensate $\langle\bar{q}q\rangle^T$ (normalized to its vacuum value) together with that of the Polyakov loop expectation value $\Phi$. In the nonlocal model without Polyakov loop the critical temperature for the chiral restoration is $T_c=116$ MeV, whereas the pure gauge sector has a critical temperature for deconfinement $T_d=270$ MeV, fixed from lattice data for $\Phi$. When coupling the quark and gluon sectors, these critical temperatures get synchronized so that $T_c\approx T_d \approx 200$ MeV at the MF level.

Near $T_c$ there is a wiggle in the behavior of the quark condensate which is caused by the perturbative nature of $1/N_c$ expansion scheme. Namely, in the region of temperatures $183$--$223$ MeV the correction to the quark condensate is larger than $1/N_c$ which one can naively expect. So, it seems reasonable to use some interpolation between the stable regions at low and high temperatures.

On the other hand, $1/N_c$ corrections slightly lower the temperature of the chiral phase transition in comparison with the mean-field result. For low temperatures $\leq 100$ MeV our result for the quark condensate practically coincides with the $\chi PT$ result, whereas the high T region is well controlled by the mean field. Near the phase transition the \textit{perturbative} $1/N_c$ expansion breaks down and the prediction of the strict $1/N_c$ expansion scheme seems not reliable in a region of $\pm 20$~MeV around the phase transition point.

\end{document}